# A Dye-Sensitized CdSe Nanocrystal Optical Transistor with High ON/OFF Ratio in the First Telecom Window with 74 ns Rise Time

*Krishan Kumar[1], Quan Liu[1,2], Jonas Hiller[1], Christine Schedel[1], Alfred Meixner[1,3], Kai Braun[1], Jannika Lauth[4,5], Marcus Scheele[1,3,*]*

1 Institute for Physical and Theoretical Chemistry, University of Tübingen, Auf der Morgenstelle 18, 72076 Tübingen, Germany.

2 Charles Delaunay Institute, CNRS Light, nanomaterials, nanotechnologies (L2n, former "LNIO") University of Technology of Troyes, 12 rue Marie Curie - CS 42060, 10004 Troyes Cedex, France

3 Center for Light-Matter Interaction, Sensors & Analytics LISA+, University of Tübingen, Auf der Morgenstelle 15, 72076 Tübingen, Germany

4 Institute for Physical Chemistry and Electrochemistry, Universität Hannover, Callinstr. 3A, 30167, Hannover, Germany

5 Cluster of Excellence PhoenixD (Photonics, Optics, and Engineering – Innovation Across Disciplines), Hannover, Germany.

**Abstract**

We report an optically gated transistor composed of CdSe nanocrystals (NCs), sensitized with the dye Zinc β-tetraaminophthalocyanine for operation in the first telecom window. This device shows a high ON/OFF ratio of six orders of magnitude in the red spectral region and an unprecedented 4.5 orders of magnitude at 847 nm. By transient absorption spectroscopy, we reveal that this unexpected infrared sensitivity is due to electron transfer from the dye to the CdSe NCs within 5 ps. We show by time-resolved photocurrent measurements that this enables fast rise times during near-infrared optical gating of $74 \pm 11$ ns. Electronic coupling and accelerated non-radiative recombination of charge carriers at the interface between the dye and the CdSe NCs are further corroborated by steady-state and time-resolved photoluminescence measurements. Field-effect transistor measurements indicate that the increase in photocurrent upon laser illumination is mainly due to the increase in carrier concentration while the mobility remains unchanged. Our results illustrate that organic dyes as ligands for NCs invoke new optoelectronic functionalities, such as fast optical gating at sub-bandgap optical excitation energies.

**Introduction**

Optical transistors are key components in optical data communication, where they convert an incoming pulse of optical information into an electrical data output.[1,2] Light pulses in the first telecommunication window (800-900 nm) are useful for specialized communication systems over short distances where the rather large dispersion at these wavelengths is irrelevant.[3] The band edge absorption and excellent compatibility with standard CMOS technology of silicon are in principal well-suited for it to serve as the active material in an optical transistor, however its indirect bandgap and intrinsically limited optical sensitivity have moved other materials, such as GaAs,

SiGe or graphene, into the spotlight.[4–6] Inorganic semiconductor nanocrystals (NCs) are considered as alternative materials for optical transistors due to their exceptionally large extinction coefficients and absorption cross-sections.[7,8] CdSe is the technologically most mature example for this material class, and electro-optical conversion with CdSe NCs has been studied for over two decades.[9,10] Tailoring the surface chemistry has mitigated the effect of frequent surface defects, crystal grain boundaries and barriers to charge carrier injection in thin films of CdSe NCs, enabling charge carrier mobilities which are comparable with those in silicon.[11,12] A large tolerance for a wide range of substrates, including flexible and bendable materials, as well as photolithographic techniques to pattern CdSe NC-based devices with high fidelity render this material class increasingly competitive with established bulk inorganic semiconductors for electro-optical application.[13,14] However, the band edge absorption in bulk CdSe is limited to < 730 nm, which is reduced further to <650 nm for typical CdSe NCs due to quantum confinement. Below these wavelengths, CdSe NCs have demonstrated excellent optical transistor properties with ON/OFF ratios of 6 orders of magnitude and detectivities > $10^{13}$ Jones.[15,16] However, due to the lack of absorbance at telecommunication wavelengths, the application of CdSe NCs in optical transistors is unattractive. This hampers the exploration of electro-optical communication units ("optical transceivers") based entirely on CdSe NCs despite their otherwise attractive optoelectronic performance. A possible solution for this shortcoming is the design of hybrid materials, for instance by mixing CdSe NCs with graphene, black phosphorous or transition metal dichalcogenides.[17–19] However, the relatively long lifetimes of separated charges at the interface of these hybrid materials invoke rise and fall times between 60 ms – 2.8 s, which so far prevents fast data communication.

Here, we demonstrate how tethering the organic dye Zinc β-tetraaminophthalocyanine (Zn4APc) to the surface of iodide-capped CdSe NCs expands the electro-optical response of thin films of this hybrid material into the first telecommunication window. At 847 nm photoexcitation, we obtain an ON/OFF ratio of 4.5 orders of magnitude and a rise time < 10 ms. With transient absorption spectroscopy (TAS), we show that the mechanism behind this action is a photoexcitation of singlet excitons in the dye, followed by charge carrier separation and electron transfer onto the n-type NCs. Elevated electron mobilities within the network of iodide-capped CdSe NCs enable fast transfer of the photoexcited charge carriers to the terminals of an optical transistor, which merely consists of two contacts, a light source and a thin layer of the hybrid nanomaterial. This work details how modifying the ligand sphere of NCs with organic π-systems generates new optical properties without compromising the electronic performance of the material. This provides attractive application perspectives, for instance for optical communication technologies.

**Results**

*Characterization of Ligand Exchange and effect on electric transconductance*

Ligand exchange of 5.5 nm wurtzite CdSe NCs is monitored by optical absorption and Raman spectroscopy of thin, solid-state films in **Figure 1a+b**. As synthesized CdSe NCs capped with a mixture of hexadecylamine (HDA) and oleic acid (OA) exhibit an excitonic transition at 621 nm (**Fig. 1a, red curve**). Surface modification with NH$_4$I (**Fig. 1a, blue curve**), followed by further ligand exchange with Zn4APc (**Fig. 1a, orange curve**) invokes a red shift of the excitonic transition by 2-3 nm and 10 nm, respectively. We attribute this either to the change in the dielectric environment of the NCs or possibly to reduced quantum confinement due to improved interparticle coupling.[20,21] The appearance of two additional absorption bands at 710 nm and 850 nm after ligand exchange with Zn4APc is supporting evidence for a successful surface modification of the CdSe NCs.[22] The Raman spectra in **Figure 1b** further corroborate this: HDA/OA capped CdSe NCs (**red curve**) are weakly Raman active and mainly show the bands of the silicon substrates at 520 cm$^{-1}$ and 940-985 cm$^{-1}$. After surface modification with NH$_4$I (**blue curve**), we observe strong bands at 207 cm$^{-1}$, 415 cm$^{-1}$ and 621 cm$^{-1}$ which we interpret as the longitudinal optical phonon (1LO) and overtone (2LO, 3LO) modes of CdSe in reasonable agreement with previous reports.[23,24] In addition to these bands, the Raman spectrum of the CdSe/I$^-$/Zn4APc film exhibits additional bands beyond 688 cm$^{-1}$, which we attribute to the presence of Zn4APc upon comparison with the spectrum of pure Zn4APc (**black curve**).[25]

Sayevich *et al.* have shown that initially insulating CdSe NCs capped with HDA/OA can be converted into n-type field-effect transistors upon surface modification with NH$_4$I and ON/OFF ratios of 5 orders of magnitude.[26] Our transconductance measurements of the CdSe/I- NC thin films in **Figure 1c** confirm these properties. While we were not successful in fabricating similar transistors by direct ligand exchange of HDA/OA-capped CdSe NCs with Zn4APc, we find in **Figure 1d** that a consecutive surface modification into CdSe/I$^-$/Zn4APc films leads to n-type transistors and an ON/OFF-ratio of 4 orders of magnitude. A comparison of **Figure 1c** with **1d** illustrates that the slightly inferior ON-OFF ratio after Zn4APc-modification is mainly due to a larger threshold voltage and hysteresis. We anticipate that charge carrier transport in CdSe/I$^-$/Zn4APc films most likely manifests without the aid of molecular orbitals of Zn4APc. In this scenario, the increased hysteresis in the hybrid film probably originates from burying the conductive CdSe/I$^-$ layer within a matrix of Zn4APc.

We determine the field-effect mobilities ($\mu$) in the dark and under 637 nm excitation by applying the gradual channel approximation and extract the charge carrier concentration ($n$). On average, we find $\mu = 2.8 * 10^{-3}$ cm$^2$/Vs and $n = 3.0 * 10^{11}$ cm$^{-3}$ in the dark which compares to $\mu = 5.2 * 10^{-3}$ cm$^2$/Vs and $n = 2.4 * 10^{16}$ cm$^{-3}$ under 35 µW incident optical power.

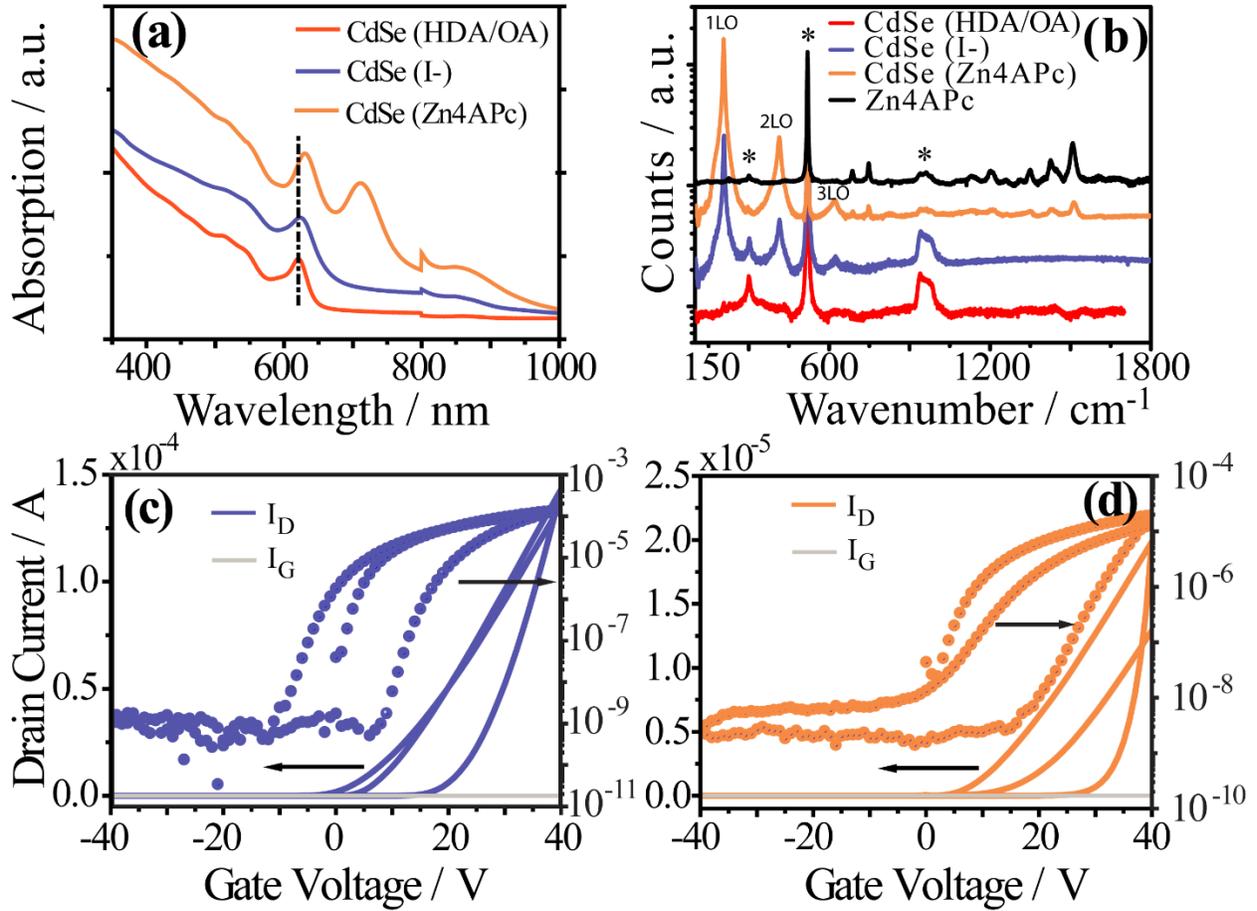

**Figure 1** (**a**) Absorption profile of thin films on glass substrates of CdSe NCs with different capping ligands as specified in the legend (**b**) Raman spectra on Si/SiO$_x$ substrates of CdSe NCs with different capping ligands as specified in the legend as well as pure Zn4APc ligand (black). Raman peaks of the Si substrate are indicated by asterisks. (**c**) Transfer characteristics of a field-effect transistor with a thin film of CdSe/I$^-$ NCs and (**d**) with CdSe/I$^-$/Zn4APc. The dotted data in (c,d) are shown on a logarithmic scale, while data plotted with solid lines is displayed on a linear scale. Grey lines represent the gate leakage in both devices.

*Optical Transistor*

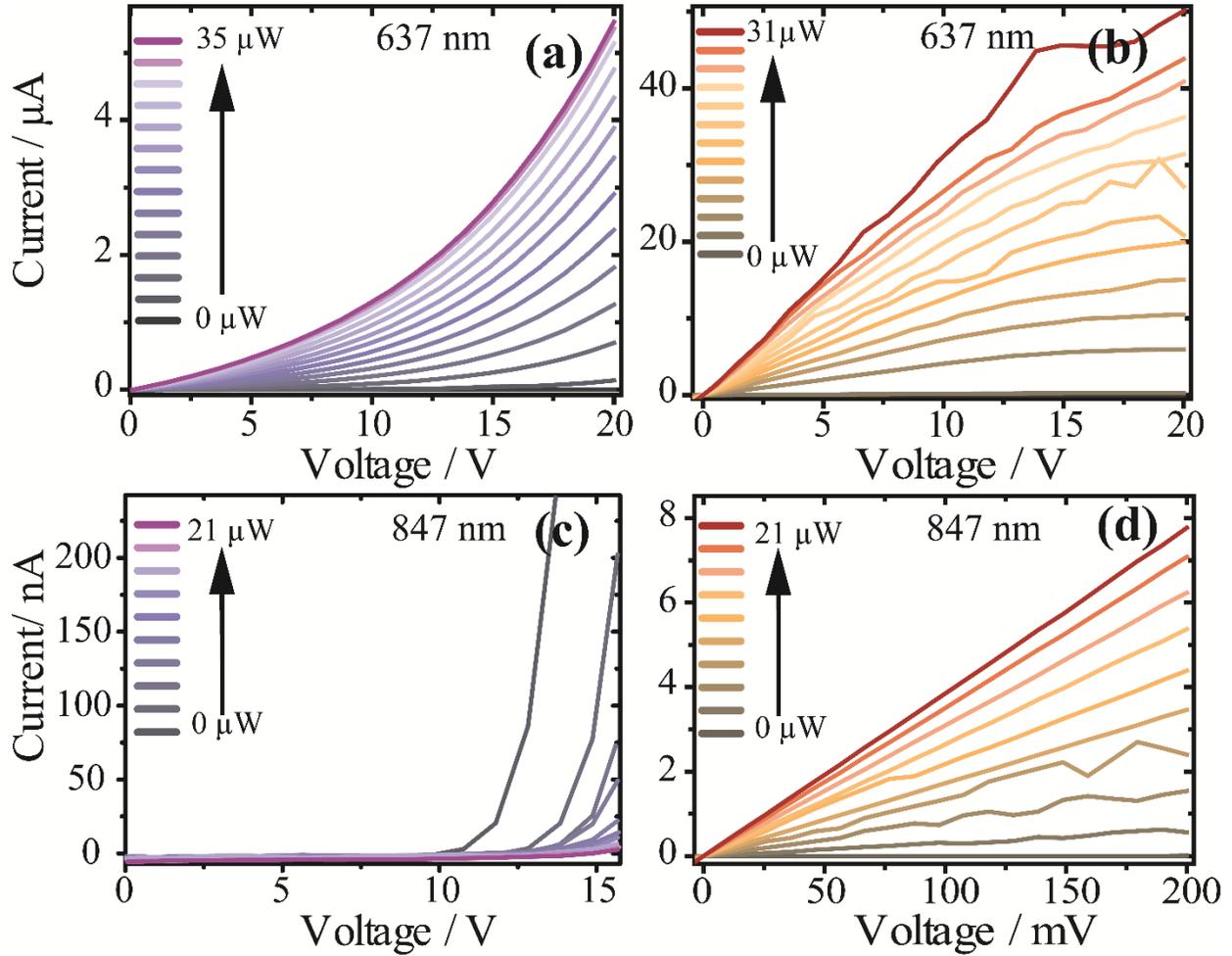

**Figure 2** Optical gating of **a)** CdSe/I⁻ NCs at 637 nm and an incident optical power of 0-35 μW, **b)** CdSe/I⁻/Zn4APc NCs with 637 nm and 0-31 μW, **c)** CdSe/I⁻ NCs with 847 nm and 0-21 μW as well as **d)** CdSe/I⁻/Zn4APc NCs with 847 nm and 0-21 μW.

We hypothesized that rather than in a *field-effect* transistor where the concentration of free charge carriers is modulated through a thin dielectric, the hybrid CdSe/I⁻/Zn4APc NC film may be more promising for application in an *optical* transistor. In such a device, charge carrier modulation is provided by an optical gate, that is, the photoexcitation by an external light source. **Figure 2** compares the performance of the CdSe/I$^-$ NC films (**Figure 2a+c**) with CdSe/I$^-$/Zn4APc NC films (**Figure 2b+d**) under optical gating with 637 nm and 847 nm, respectively. Under near-resonant excitation of the CdSe NCs with 637 nm (**Fig. 2a+b**), we observe strong optical modulation of the current output for both materials. Without Zn4APc, the I/V-characteristics is mostly still in the linear regime (**Fig. 2a**), while after additional surface-functionalization with the phathalocyanine, the ON-currents are one order of magnitude higher at otherwise same excitation power and the I/V-characteristics approach the saturation regime (**Fig. 2b**). We argue that this may be the result of the additional absorption of Zn4APc at 637 nm, resulting in a larger photocurrent and a more efficient photogeneration of free charge carriers.

A vastly different behaviour of CdSe/I$^-$ *vs.* CdSe/I$^-$/Zn4APc is observed under optical gating with 847 nm (**Fig. 2c+d**). CdSe NCs show very weak absorption by in-gap defect states at this wavelength and consequently, the expected photocurrent is small. Below a source-drain voltage ($V_{SD}$) of 10 V, there is no clear trend of the photocurrent with increasing excitation power, and the ON-current indeed largely equals the OFF-current (**Fig. 2c**). For large electric fields, e.g. $V_{SD}$ > 10 V, the I/V-characteristics show non-ohmic behaviour and an OFF-current exceeding the ON-current, that is, a negative photoeffect. Such an effect is sometimes observed at sub-bandgap excitation of semiconductors with a significant number of shallow in-gap states.[27] Briefly, for an n-type semiconductor with in-gap states near the conduction band edge, the OFF- or dark-current results from free electrons in the conduction band donated by the shallow in-gap state. Under sub-bandgap optical excitation, electrons are excited from the valence band edge into the (empty) in-gap state. This leaves trapped electrons in the in-gap states, free electrons in the conduction band and free holes in the valence band. If fast recombination of the free electron/hole pair is possible, the ON-current in this material will be smaller than the OFF-current, resulting in a seemingly negative photoeffect as shown in **Fig. 2c**. In contrast, after functionalization with Zn4APc the photoeffect under 847 nm excitation is strongly positive (**Figure 2d**). At this wavelength, only the phthalocyanine shows considerable absorption, and we suggest that Zn4APc acts as a sensitizer to activate the CdSe/I$^-$ network for photoconduction and operation as an optical transistor.

*Wavelength-dependent ON/OFF properties*

A key property of an optical transistor is its ability to distinguish between a poorly conductive OFF-state and a highly conductive ON-state. In addition, it is desirable to maintain a large ON/OFF-ratio over a wide wavelength range to achieve a spectrally broad photosensitivity. In **Figure 3**, we investigate the ON/OFF properties of CdSe/I$^-$ NCs with (**orange curve**) and without (**blue curve**) sensitization with Zn4APc under optical gating with 637 nm (**Fig. 3a**) as well as 847 nm (**Fig. 3b**) at a bias of 200 mV. At 637 nm and an incident optical power of 35 µW, the Zn4APc-sensitized CdSe/I$^-$ film outperforms the same NCs without the organic π-system with an ON/OFF-ration of 6 orders of magnitude *vs.* 4.6 orders of magnitude. At 847 nm and 31 µW absorbed optical power, the sensitizing effect of Zn4APc exhibits the highest impact with an ON/OFF ratio of 4.5 orders of magnitude *vs.* < 2 orders of magnitude for the CdSe/I$^-$ NC film only. We note again that the small photosensitivity of the CdSe/I$^-$ NCs is most likely due to in-gap states because of surface defects, which could be reduced even further by better passivation. This view is supported by the higher OFF-current after the first optical excitation cycle, which indicates residual charging of the film. Moreover, the observation of a negative photoeffect (**Figure 2c**) is typical for the presence of in-gap states. The ON-OFF characteristics in **Figure 3** are in good agreement with the optical gating measurements in **Figure 2** in that sensitization with Zn4APc invokes improved optical switching at 637 nm and especially 847 nm by additional 2.5 orders of magnitude. We note that the $10^6$ ON/OFF ratio at 637 nm is amongst the highest for this material, which is otherwise only achieved for single nanowires and rather sophisticated device architectures.[15,16,28–31] The large sensitivity at 847 nm is unprecedented for CdSe NCs due to its large bandgap and usually requires alloying, e.g. with CdTe, to achieve a significant absorption in the near-infrared.[32]

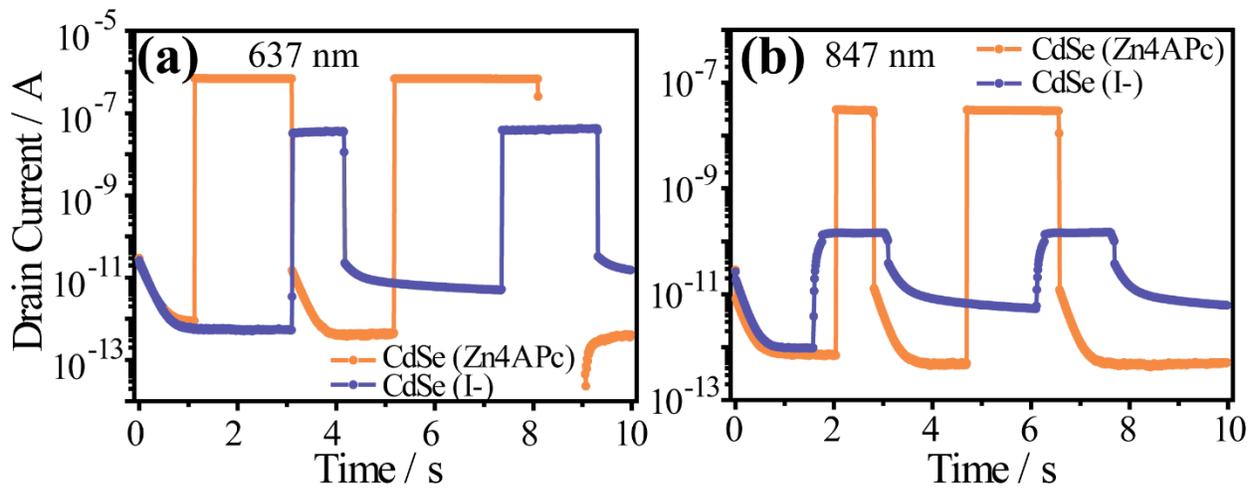

**Figure 3**. ON/OFF Properties of optical transistors at $V_{sd}$ = 0.2 V of thin films of CdSe/I$^-$ NCs (blue) and CdSe/I$^-$/Zn4APc NCs (orange). In **a)**, the excitation source is provided by 35 μW of 637 nm incident light and in **b)** by 21 μW of 847 nm light. The time resolution in both experiments is 10 ms per step, which is the integration time of the current measurement unit.

*Transient Absorption Spectroscopy (TAS)*

To understand the sensitization mechanism exerted by the organic π-system onto the NCs, we study optically thick films of CdSe/I$^-$ cross-linked with Zn4APc by TAS. The samples are photoexcited with short laser pulses (~180 fs) of different wavelengths (640 – 850 nm, 1.9 – 1.5 eV), and the differential change in absorption is examined by broadband probe pulses between 500 and 900 nm.[22,33] The color-coded 2D TA spectrum obtained under near-resonant excitation of the CdSe NCs (pump at 640 nm or 1.94 eV) is displayed in **Figure 4a** and exhibits five bands, labelled 1, 1a, 2, 3 and 4. We attribute the weak bands 1 and 1a (710-780 nm or 1.75-1.59 eV) to the convoluted bleach of the Davydov-split HOMO-LUMO transition of Zn4APc and note that such splitting is often observed for aggregates of these molecules.[22,34–36] We assign the strong bleach in band 2 (625 nm or 1.96 eV) to the $1S_h$-$1S_e$ transition of the CdSe NCs. Similarly, the induced absorption band 3 (580 nm or 2.14 eV) and the bleach in band 4 (520 nm or 2.38 eV) most likely originate from a biexcitonic shift and the $1P_h$-$1P_e$ transition in the CdSe NCs as detailed previosuly.[37]

The TA spectrum changes significantly under near-resonant excitation of the HOMO-LUMO transition of Zn4APc (pump at 800 nm or 1.55 eV, **Figure 4b**). Bands 1 and 1a are now very prominent, supporting our assignment as the Davydov-split Q-band (the $S_0 \rightarrow S_{11}$ and $S_0 \rightarrow S_{12}$ transitions) of Zn4APc.[34] This is further detailed by the normalized line cuts at delay times $\geq$ 7 ps in **Figure 4c** (black line: pump at 640 nm, red line: pump at 800 nm). Most importantly, we observe bands 2 and 3, although a direct excitation of these CdSe-related transitions is not possible at 800 nm. We hypothesize that a transfer of photoexcited charges from the π-system onto the NCs is responsible for this finding. To test this hypothesis, we analyze the decay of the transient bands 1/1a and 2 towards a possible time correlation in **Figure. 4d**. Indeed, we find the *decay time* of the bleach in band 1 matching the *rise time* of the bleach feature in band 2 at early times of the measurement (25 ps), before bleach 2 starts decaying again. The decay dynamics of the Zn4APc-associated bleach feature 1 and the correlated indirect bleaching of the CdSe feature 2 are fitted with a double-exponential ($\tau_1 = 330 \pm 25$ fs, $\tau_2 = 5.8 \pm 0.2$ ps and $\tau_1 = 300 \pm 34$ fs, $\tau_2 = 4.7 \pm 0.1$ ps, see **Fig. 4d,f**) and almost the same time constants. Note that when studying the decay dynamics of bleaches in band 1/1a and 2 at a direct photoexcitation of the CdSe at 640 nm, bleach feature 2 rises instantly (see **Fig. S1** and for further details **Fig. S2** and **S3**).

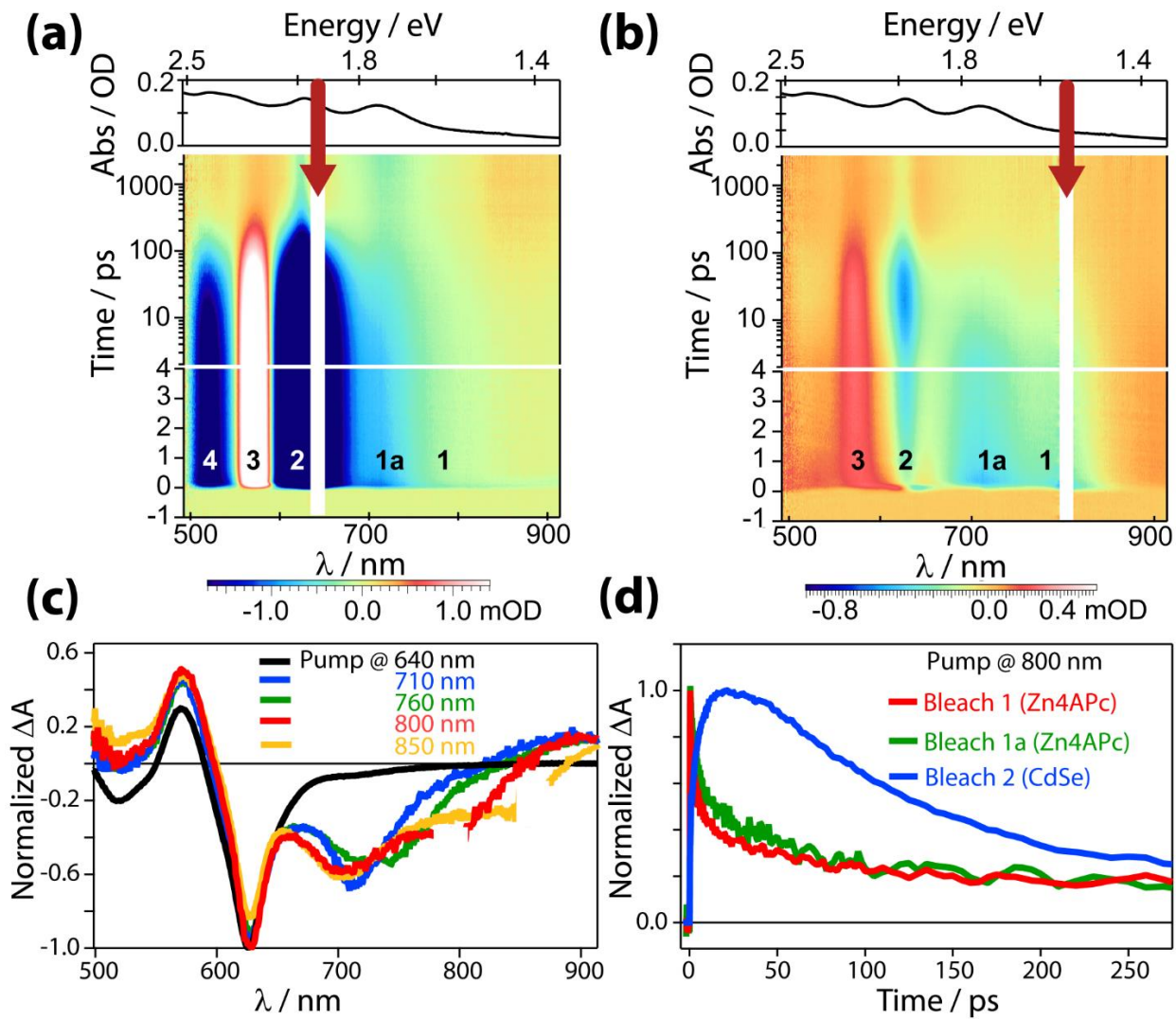

**Figure 4**. 2D-TA spectra of a CdSe/I⁻/Zn4APc film. (**a**) Near-resonant direct photoexcitation of CdSe NCs at 640 nm mainly leads to bleach features from the CdSe NCs and only weak contribution from the Zn4APc. (**b**) Photoexcitation near the Zn4APc optical transition at 800 nm leads to charge transfer from the Zn4APc to the CdSe NCs visible by an indirect bleach of the CdSe transitions and visible by the rise of bleach feature 2 with the same time constants as the decay of bleach feature 1 as discussed in the text. **(c)** Spectral slices of the CdSe/I⁻/Zn4APc film photoexcited at different wavelengths and associated increased contribution of the Zn4APc molecule at higher wavelengths and considerable Zn4APc absorption. **(d)** Temporal slices of band 1 (red, 785 nm), band 1a (green, 710 nm) and band 2 (blue, 621 nm) with ultrafast rise in the first 25 ps of the measurement. The red and the blue curve exhibit almost equal time constants as discussed in the text.

*Time-resolved photocurrent measurements*

A crucial figure of merit of an optical transistor is its rise time under optical excitation with a square pulse signal. The typical time-resolved response of a CdSe/I⁻-Zn4APc based device with an active area of 2.5 µm × 1000 µm on Si/SiO$_2$ is presented in **Figure 5**. We first measure the impulse response of the device towards 636 nm (**Figure 5a**) delta function laser pulses (< 500 ps pulse length, 3 MHz repetition rate) under varying bias of 0.2 – 10 V. Trapezoidal integration of this impulse response in the time domain yields the first half of the corresponding square pulse response and allows determining the rise time. (**Figure 5b**, red lines.) To verify this evaluation, we also measure the time-resolved photocurrent of the same device to a 635 nm square pulse with 100 ns pulse length and 3 MHz repetition rate in **Figure 5b** (gray lines). We find excellent agreement in the onset of the time-resolved photocurrent obtained with these two independent measurements for all applied biases. This suggests that integration of the impulse response to a 779 nm delta pulse (**Figure 5c**, < 500 ps pulse length, 3 MHz repetition rate) will also allow a reliable measurement of the rise time. The rise time, defined as the time elapsed for the integrated photocurrent to rise from 10 % to 90 % of its maximum value, at 0.2 V is 45 ± 20 ns under 636 nm laser excitation and 74 ± 11 ns under 779 nm laser illumination. Increasing the applied bias to 10 V not only increases the photocurrent but also the rise time to 128 ± 80 ns and 153 ± 51 ns under 636 and 779 nm laser excitation, respectively. While the former is the expected effect of an increased electric field, the latter indicates the population and/or formation of deep traps under high field conditions.

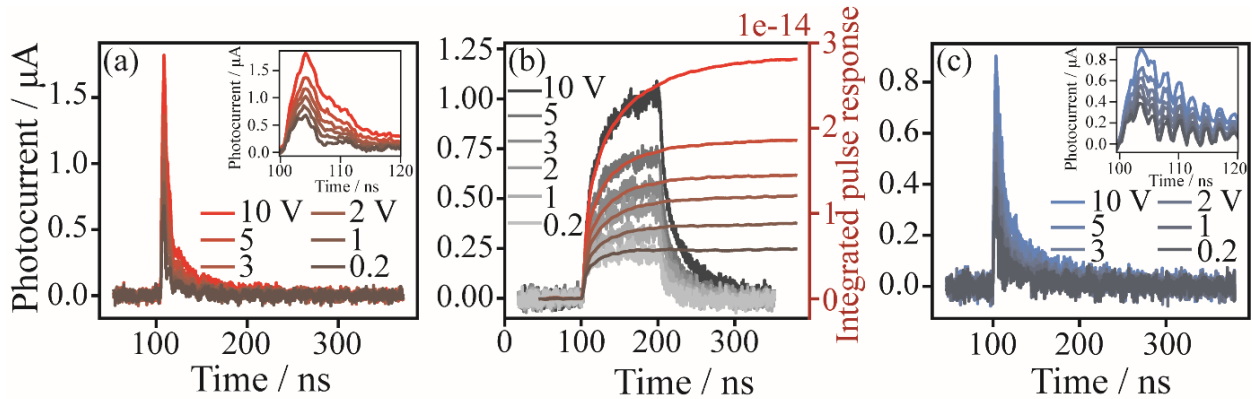

**Figure 5. (a)** Impulse photocurrent response of CdSe/I⁻-Zn4APc thin films towards 636 nm delta function pulses (< 500 ps), 600 µW optical power, 3 MHz repetition rate and varying bias of 0.2 – 10 V. **(b)** Trapezoidal integration of the impulse response in (a) (red lines) as well as the measured photocurrent response to a square pulsed 635 nm excitation with 100 ns pulse length (< 12 mW optical power, 3 MHz) and varying bias of 0.2 – 10 V (grey lines). **(c)** Impulse photocurrent response of the same device towards 779 nm delta function pulses (< 500 ps, 600 µW, 3 MHz) and varying bias of 0.2 – 10 V.

*Fluorescence (Lifetime) measurements*

To investigate the radiative excited-state decay of the CdSe/I⁻-Zn4APc films, we perform fluorescence lifetime measurements upon excitation with 488 nm in **Figure 6** (orange). For comparison, we also display the fluorescence properties of thin films of CdSe/I⁻ NCs without Zn4APc (blue). Since we observe that the fluorescence spectra and particularly the decay kinetics exhibit a substantial dependence on the substrate coverage and film thickness, we depict the same measurements for sub-monolayers (**Figure 6a+b**) as well as multilayers (roughly four layers, **Figure 6c+d**). For sub-monolayer coverage, the CdSe/I⁻ NC film displays a single fluorescence band with a maximum at 617 nm, while the Zn4APc-capped film shows a similar signal at 631 nm and additional broad bands at 731 nm as well as 570 nm (**Figure 6a**). We assign the bands at 617 nm and 631 nm, respectively, to the (red-shifted) band edge fluorescence of the NCs and tentatively attribute the 731 nm signal to the singlet fluorescence of the organic π-system. For multilayers of the CdSe/I⁻/Zn4APc films, we find peaks at 657 nm and 745 nm, which we attribute again to the further redshifted CdSe band edge and Zn4APc singlet fluorescence, respectively (**Figure 6c**). The third band at 570 nm is still visible but substantially weakened. Multilayers of CdSe/I⁻ NCs show the same steady-state fluorescence as sub-monolayers.

For time-resolved fluorescence decay measurements, the fluorescence signal is integrated over the full visible regime and therefore contains contributions from the NCs as well as the organic π-system. For sub-monolayers (**Figure 6b**), all decay curves are well-fitted with biexponentials, exhibiting a slow and fast decay component with time constants $\tau_1$ and $\tau_2$, respectively. For CdSe/I⁻, we observe $\tau_1$ = 6.1 ns and $\tau_2$ = 1.4 ns. After cross-linking with Zn4APc, the radiative decay is significantly faster with $\tau_1$ = 3.1 ns and $\tau_2$ = 0.62 ns. The time constants for the CdSe/I⁻ NCs are consistent with earlier results on single CdSe/CdS NCs passivated with inorganic ions in the solid state, but significantly smaller than those obtained for CdSe/I⁻ NCs in solution.[38,39] For multi-layers, the fluorescence decay is substantially faster (**Figure 6d**): Without Zn4APc, we now find $\tau_1$ = 5.3 ns and $\tau_2$ = 0.86 ns, while for Zn4APc tethering, the decay is too fast to be distinguished from the instrument response function of our measurement set-up, such that no kinetic parameters can be extracted.

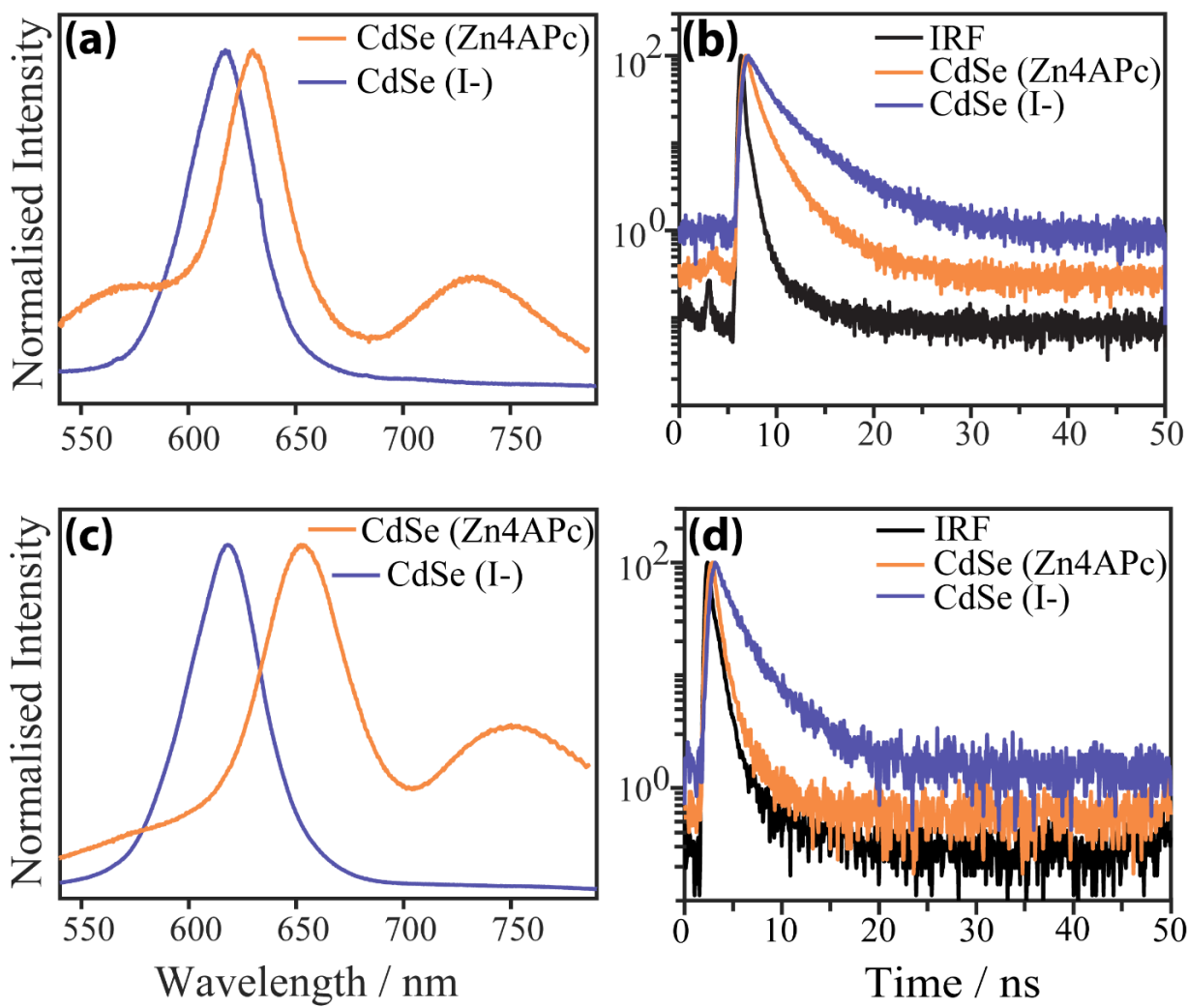

**Figure 6. a)** Steady-state fluorescence of a (sub-)monolayer of CdSe/I⁻ NCs (purple) and CdSe/I⁻ NCs capped with Zn4APc (orange). **b)** Time-resolved fluorescence decay of a similar (sub-)monolayer of CdSe/I⁻ NCs without (purple) and with (orange) Zn4APc. The instrument response-function is depicted in black. **c)+d)** Steady-state and time-resolved fluorescence decay of multilayers of the same two materials. In all cases, the excitation wavelength is 488 nm.

**Discussion**

CdSe NCs are well-known for their high performance in field-effect transistors with low threshold voltage, fast switching times, printability and lithographic processability.[12,14,40,41] Our results suggest that they are also excellently suited for application as optical transistors.[16,42,43] Large extinction coefficients ($\varepsilon = 2.7 * 10^5$ L/mol)[44] and small intrinsic carrier concentrations on the order of $10^{11}$ cm$^{-3}$ invoke high ON/OFF ratios of up to six orders of magnitude for optical transistors with a relatively simple light source. Particularly noteworthy is the ON/OFF ratio of 4.5 orders of magnitude at 847 nm illumination, which is not only unprecedented for CdSe, but also challenging to realize with other intrinsic IR absorbers, such as lead or mercury chalcogenides. Typical ON/OFF ratios for PbSe or PbS NCs are in the range of three orders of magnitude and often require additional electric gating because of the higher intrinsic carrier concentrations ($10^{16}$ cm$^{-3}$).[33,45,46] For HgTe quantum wells with an absorption at 900 nm and an intrinsic carrier concentration of $10^{12}$ cm$^{-3}$, an ON/OFF ratio of 2 orders of magnitude at room temperature was reported.[47] We explain the favorable performance of Zn4APc-functionalized CdSe NCs in this regard with a sensitization of the NCs by the organic dye ($\varepsilon$(711 nm) $\approx 3.6 * 10^4$ L/mol)[48], enabling near infrared absorption and harvesting of additional photons. Electronically, the dye marginally reduces the dark current and increases the threshold voltage (compare **Figure 1c/d**) towards electric gating. This indicates that charge carrier transport proceeds mostly across a conductive network of iodide-capped NCs, including the carriers which were originally photoexcited in the organic dye. The largely identical field-effect mobilities extracted in the dark and under full optical excitation (0.0028 *vs.* 0.0052 cm$^2$/Vs) demonstrate that photoexcited charge carriers use the same transport channels as in the dark. This is in contradiction to photocurrent measurements on PbS and PbSe NC films, which displayed increased mobilities under photoexcitation.[45,49,50] We note, however, that the excitation power densities in those studies were substantially smaller than in the present case (50 μW/cm$^2$ vs. ~100 mW/cm$^2$).

The benefit of the dye-sensitization approach presented here is further illustrated by the favorable time-resolved response to optical gating. While the rise time of the pure CdSe/I⁻ film with 160 ms is slow at sub-bandgap excitation (**Figure 3b**), optical transistors made of Zn4APc-sensitized CdSe/I⁻ NCs are over one million times faster with 74 $\pm$ 11 ns under 779 nm excitation (**Figure 5c**). Other materials with strong absorption in this spectral regime, such as HgTe-CdS nanoplatelets or PbSe NC nanojunctions, have shown comparably slower rise times of 10 µs and 300 ns, respectively.[51,52] Recent time-resolved photocurrent measurements on lead halide perovskite photodetectors have reported response times as fast as 1.8 ns, but these were limited to excitation at 532 nm.[53] We suggest that the dye-sensitization concept outlined here may allow expanding the spectral window for optical gating with perovskites similar to what we have demonstrated with CdSe NCs to reward even faster NIR optical transistors.

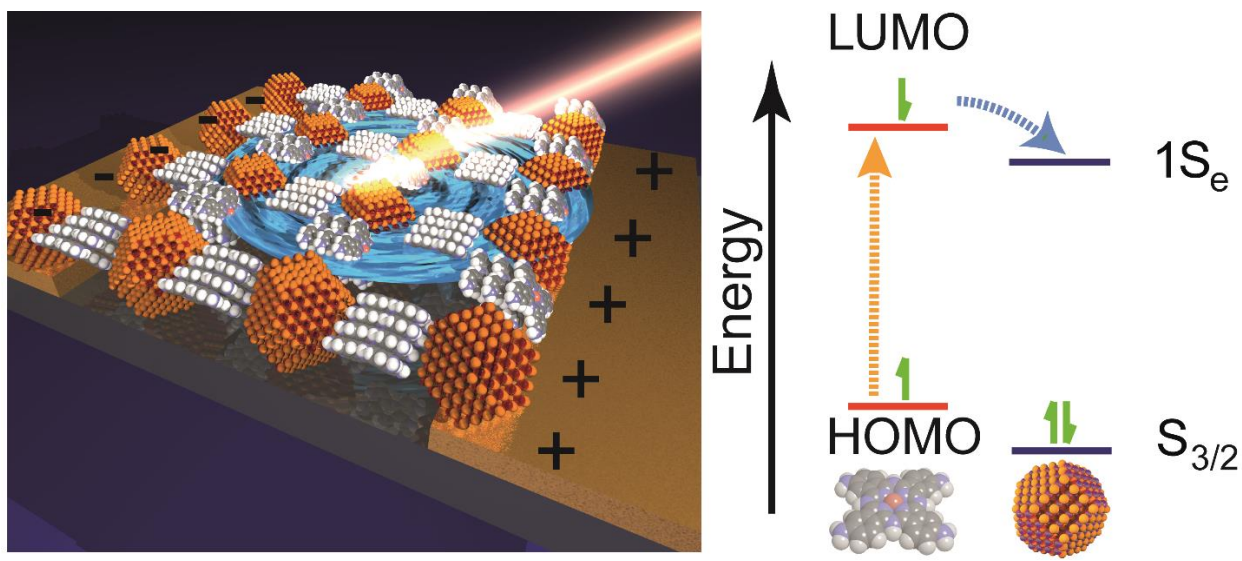

**Scheme 1.** Idealized schematics of the structural composition and working mechanism of the optical switch under near-IR excitation. Direct photoexcitation of the HOMO-LUMO transition of the organic dye is followed by electron transfer into the $1S_e$ state of the NCs, from where the electrons are shuttled to the electrodes under a small bias.

At the heart of this concept is charge transfer from the dye to the NCs, which is evident from the transient absorption data in **Figure 4b+c** at various excitation wavelengths between 710 – 850 nm. The essentially identical time constants for the biexponential decay of the photobleach of the singlet transition of Zn4APc compared to the biexponential rise of the excitonic photobleach of CdSe are a strong hint for such a mechanism (**Figure 4d**). Considering the slower decay/rise component as the rate limiting step of charge transfer ($\tau \approx 5$ ps) illustrates that optical switching at this hybrid interface is many orders of magnitude faster than in other hybrid CdSe NC based materials.[17–19] On the other hand, resonant excitation of the NCs at 640 nm exhibits no comparable signs of a charge transfer in the opposite direction (**Figure 4a**). From this, we can draw two conclusions: I) While an unambiguous distinction between charge and energy transfer is challenging, charge transfer seems to be the more plausible mechanism here. This follows from the fact that energy transfer requires a large overlap between the absorption of the acceptor and the fluorescence of the donor. Thus, energy transfer from the NCs onto the molecule should be much more favorable than in the reverse direction; however, we only observe the latter. II) The frontier orbital energies of Zn4APc are such that electron transfer into the $1S_e$ state of CdSe is thermodynamically favorable. Electrochemical studies suggest that the LUMO of Zn4APc is indeed approx. 0.5 eV above the $1S_e$ state of CdSe, rendering electron transfer feasible.[54] In contrast, the HOMO of Zn4APc is positioned approx. 0.4 eV above the $1S_h$ state, which speaks against hole transfer towards the NCs. Overall, we therefore attribute the large photocurrent under 847 nm excitation of Zn4APc-CdSe NCs to the singlet absorption of Zn4APc, splitting of the exciton and transfer of electrons to CdSe, which is schematically detailed in **Scheme 1**. This view is also consistent with the finding that transient photobleaching of the excitonic transition in CdSe is predominantly caused by electrons.[55] Electrons are the majority carriers in films of iodide-capped CdSe NCs, such that they are swept quickly to the electrodes. Holes remain either trapped in the dye, recombine at the organic/inorganic interface with excess electrons in the NCs or are shuttled (slowly) to the electrodes via hopping between adjacent dye molecules. To this end, the time-resolved fluorescence data in **Figure 6b+d** enable valuable insights into the recombination mechanism of the charge carriers. The experimental lifetime $\tau$ is correlated with the radiative and non-radiative rates, $k_r$ and $k_{nr}$, as $\tau = \frac{1}{k_r+k_{nr}}$. Thus, the substantial decrease of $\tau$ upon functionalization of CdSe/I⁻ with Zn4APc indicates an increase of either the radiative recombination or a non-radiative pathway. Since it is difficult to imagine how any of the afore-mentioned scenarios for the fate of the holes could lead to additional *radiative* recombination, we suspect that the shortened lifetime is due to accelerated non-radiative recombination, presumably due to trapped holes at the organic/inorganic interface and their interaction with excitons in the NCs. The same scenario could also explain the redshifted fluorescence of CdSe/I⁻ NC films after surface-functionalization with Zn4APc: If electrons reside predominantly in the NCs and holes in the organic dye, a "type-II" radiative recombination at the organic-inorganic interface would be possible.[56] This emission should be redshifted compared to the pure CdSe band edge recombination, since the Zn4APc HOMO lies energetically above the CdSe $1S_h$ state. The redshifted absorption and fluorescence (**Figure 1a and Figure 6a, c**) support this view. In contradiction, a type-II fluorescence at a semiconductor interface is typically accompanied by a substantially increased radiative lifetime, since electron and hole are spatially separated. Our finding of a drastically reduced lifetime does not support this scenario. However, it is also possible that a potential increase in the radiative lifetime is accompanied by an increase of non-radiative recombination with the result of an overall shorter experimental lifetime. In this context, we note that the rise time upon resonant excitation of the dye (74 $\pm$ 11 ns) is only marginally longer compared to resonant excitation of the NCs (45 $\pm$ 20 ns). Considering that photoexcited charges in the dye need to be shuttled to the NCs before they can contribute to the photocurrent, this implies that trapping at the organic/inorganic interface in CdSe/I⁻-Zn4APc is relatively short-lived.

**Conclusion**

Iodide-capped, n-type CdSe nanocrystals are surface-functionalized with the organic π-system Zn β-tetraaminophthalocyanine and assembled into a thin-film optical transistor. Optical gating with a 637 nm light source yields excellent current modulation by six orders of magnitude, near-saturation behavior and a rise time of 45 $\pm$ 20 ns. Under near-infrared excitation, the modulation exhibits 4.5 orders of magnitude and rise times of 74 $\pm$ 11 ns, which is unprecedented for CdSe nanocrystals due to their intrinsically poor sensitivity to infrared light. We show that this is enabled by harvesting of infrared photons by the organic dye, singlet exciton splitting and electron transfer onto the nanocrystals with a lifetime of ~5 ps. Such dye-sensitized nanocrystals combine the well-developed electronic properties of CdSe nanocrystals with high optical sensitivity in the near-infrared, which is attractive for application in optical transceivers operating in the first telecommunication window.

ASSOCIATED CONTENT

**Supporting Information**.

(**S1**) Transients of CdSe/I⁻/Zn4APc after 640 nm pump pulses, (**S2**) Transients of CdSe/I⁻/Zn4APc after 800 nm pump pulses, (**S3**) Transients of CdSe/I⁻/Zn4APc after 710 nm pump pulses, (**S4**) Schematics of the time-resolved photoluminescence instrument. Experimental details and methods.

AUTHOR INFORMATION

**Corresponding Author**

*Email: marcus.scheele@uni-tuebingen.de

**Author Contributions**

The manuscript was written through contributions of all authors. All authors have given approval to the final version of the manuscript.

ACKNOWLEDGMENT

The authors acknowledge the DFG for support under Grant SCHE1905/3 and under Germany's Excellence Strategy within the Cluster of Excellence PhoenixD (EXC 2122, Project ID 390833453). The time-resolved photocurrent measurements have been funded the European Research Council (ERC) under the European Union's Horizon 2020 research and innovation program (grant agreement No 802822).

# *Supporting Information*

**Fig. S1** shows temporal slices of a CdSe/I⁻/Zn4APc-COIN optical transistor photoexcited at 640 nm near-resonantly to the first excitonic transition of the CdSe NCs and – unlike measurements with a photoexcitation wavelength of ≥ 710 nm - lacking a time-correlation between the decay of bleach feature 1 and 1a and bleach feature 2 in the first 25 ps of the measurement due to a charge transfer mechanism (no rise time in bleach feature 2). The dynamics of the Zn4APc bleaches 1/1a and the CdSe NC bleaches 2 are fitted with a double-exponential ($\tau_1 = 3.7 \pm 2.3$ ps, $\tau_2 = 74 \pm 17$ ps for the rather noisy bleach feature 1 and $\tau_1 = 1.4 \pm 0.1$ ps, $\tau_2 = 34 \pm 2.0$ ps for the nicely resolved bleach feature 1a, $\tau_1 = 4.6 \pm 0.2$ ps, $\tau_2 = 71 \pm 2.5$ ps for bleach feature 2).

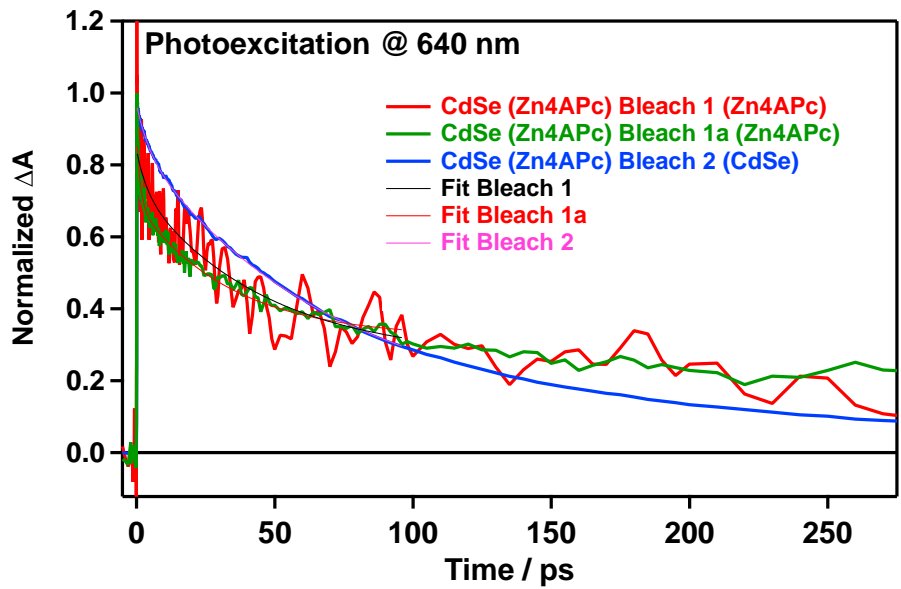

*Fig. S1. Decay of Davydov-split bleach features 1 and 1a of the Zn4APc and decay of bleach feature 2 due to direct photoexcitation of the CdSe NCs.*

By comparing the time constants of the decay in bleach feature 1 and 1a at different photoexcitation wavelengths with the rise of bleach feature 2 in the first 25 ps of the measurement, we find that at photoexcitation wavelengths of 800 nm (close to the wavelength of the photocurrent measurements), particularly photoexcited electrons from the energetically lower lying $S_{11}$ level of the Zn4APc molecule seem to be transferred to the CdSe $S_{1e}$ level, while at 710 nm, electrons from the $S_{12}$ transition in the Zn4APc molecule seem to be preferred for transfer to the CdSe $S_{1e}$ level (see **Fig. S2** and **S3** respectively). **Fig. S2** shows almost the same time constants of a biexponential fit for the decay of the directly photobleached Davydov-split $S_0 \rightarrow S_{11}$ transition (bleach feature 1, $\tau_1 = 330$ fs $\pm$ 25 fs, $\tau_2 = 5.8 \pm 0.2$ ps) photoexcited at 800 nm and the time corelated rise of bleach feature 2 due to charge transfer into the CdSe $1S_e$ level (bleach feature 2, $\tau_1 = 300 \pm 34$ fs, $\tau_2 = 4.7 \pm 0.1$ ps). Time constants for the decay of the $S_0 \rightarrow S_{12}$ transition in bleach feature 1a are $\tau_1 = 640$ fs $\pm$ 110 fs, $\tau_2 = 6.5 \pm 0.4$ ps.

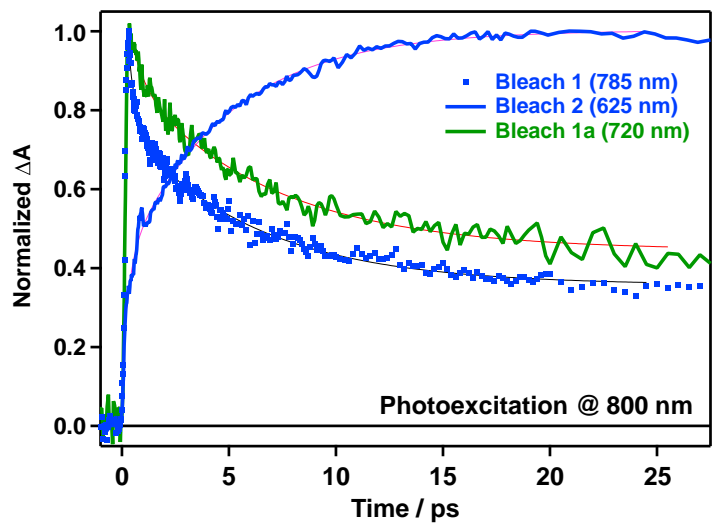

*Fig. S2. Decay of bleach feature 1 at 785 nm when photoexciting CdSe/I⁻/Zn4APc-COIN at 800 nm (Davydov-split $S_0{\rightarrow}S_{11}$ transition) and time correlated ultrafast rise (before decay at longer times) of CdSe-associated bleach feature 2.*

Photoexciting the sample at 710 nm close to the $S_0{\rightarrow}S_{12}$ transition of the Zn4APc leads to the time constants of bleach feature 1a ($\tau_1 = 550$ fs $\pm$ 26 fs, $\tau_2 = 4.8 \pm 0.3$ ps) being almost the same as in the rise of bleach feature 2 ($\tau_1 = 480 \pm 24$ fs, $\tau_2 = 3.7 \pm 0.1$ ps), whereas at 710 nm the $S_0{\rightarrow}S_{11}$ transition decays faster ($\tau_1 = 190 \pm 45$ fs, $\tau_2 = 5.0 \pm 0.5$ ps).

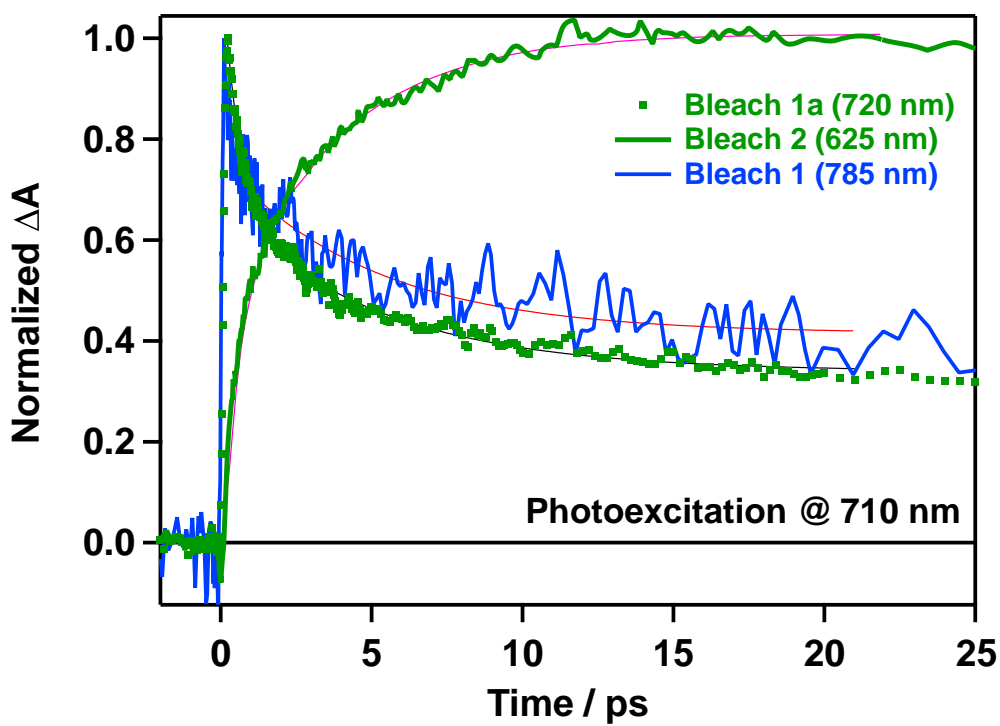

*Fig. S3. Decay of bleach feature 1a at 720 nm when photoexciting CdSe/I⁻/Zn4APc-COIN at 710 nm (Davydov-split $S_0 \rightarrow S_{12}$ transition) time correlated ultrafast rise (before decay at longer times) of CdSe-associated bleach feature 2 as discussed above.*

**Experimental Methods:**

Chemicals: Cadmium oxide (CdO, 99.99%, Aldrich), oleic acid (OA, 90%, Aldrich), trioctylphosphine (TOP, 97%, Abcr), trioctylphosphine oxide (TOPO, 99%, Aldrich), hexadecylamine (HDA, 90%, Aldrich), 1-octadecene (ODE, 90%, Acros Organics), selenium pellet (Se, 99.999%, Aldrich), ammonium iodide (99.999%, Aldrich), N-methylformamide (NMF, 99%, Aldrich), hexane (Extra Dry, 96%, Acros Organics), ethanol (Extra Dry, 99.5%, Acros Organics), acetone (Extra Dry, 99.8%, Acros Organics), dimethyl sulfoxide (DMSO, 99.7%, Acros

Organics), acetonitrile (Extra Dry, 99.9%, Acros Organics). All chemicals were stored and used inside the nitrogen-filled glovebox except the chemicals used in CdSe NCs synthesis.

**CdSe NCs synthesis**:

Wurtzite 5.5 nm CdSe NCs were synthesized using a previously reported literature procedure.[1] Briefly, 176.7 mg CdO, 8 g TOPO, 8 g HDA, 2.2 mL OA and 45.8 mL ODE were weighed into a three-neck round bottom flask and kept under vacuum for 2 h (~ $10^{-2}$ mbar). It was heated under a nitrogen atmosphere to 300 °C until the solution became clear. The solution was cooled down to 275 °C and kept for 30 min. In a glass vial, 1 M TOPSe was prepared by heating 130 mg Se in 1.6 mL TOP at 120 °C under constant stirring. Together with this TOPSe solution, 6.4 mL TOP and 8.0 mL ODE were added to the flask, and the temperature was raised to 280 °C. After that, the mixture was kept for about 25 min to grow CdSe NCs. The reaction was quenched by a sudden drop in temperature. For purification, the reaction mixture was precipitated with ethanol, centrifuged and redispersed in hexane. For further purification, this procedure was repeated twice with acetone and ethanol, and the precipitate redispersed in toluene in both cases. Finally, the NCs were purified by adding methanol and redispersion in hexane.

**Ligand exchange**:

The ligand exchange procedure was adopted from literature and performed in a nitrogen-filled glovebox.[1,2] Briefly, 300 µL of 1 M $NH_4I$ solution and 2.7 mL acetone were added to 5 mL of CdSe NCs (conc. ~10 mg/mL) in hexane and stirred continuously until the hexane layer became completely colorless. The aggregate was centrifuged, washed with hexane and centrifuged again. The precipitate was dissolved in 3 mL of NMF and centrifuged using a hexane/acetone (1/2) mixture and finally dissolved in NMF. NCs prepared in this way are referred to in the manuscript as CdSe/I⁻ NCs.

**Device preparation**:

Device fabrication was performed in a nitrogen filled glove box. A commercially available bottom-gate, bottom-contact transistor substrate (n-doped silicon (n = 3 x 10¹⁷ cm⁻³ with 230 nm thermal oxide, Fraunhofer Institute for Photonic Microsystems, Dresden, Germany) with interdigitated Au electrodes of 10 mm width and varying channel lengths (2.5 µm, 5 µm, 10 µm and 20 µm) was used for the preparation of the FETs. In a typical film preparation, CdSe/I⁻ NCs (~60-100 mg/mL) were spin coated onto the substrate at 35 rps and dried at 80 rps. The film was annealed at 190 °C for 30 min.

For devices composed of CdSe/I⁻/Zn4APc films, a solution of CdSe/I⁻ NCs in NMF was deposited onto the FET substrate together with ~30 µL of a saturated Zn4APc solution in DMSO. The mixture was left undisturbed for a sufficient amount of time to react to form a CdSe/I⁻/Zn4APc film, after which the remaining solvent was spun-off the substrate to leave a continuous film. The as-prepared film was washed with acetonitrile to remove excess and unbound Zn4APc. Finally, the film was annealed at 190 °C for 30 min.

**Electrical and Optical Measurements**:

Electrical measurements were carried out under nitrogen by using a *Keithley 2634B Source Meter*. The charge carrier mobility (µ) was extracted using the gradual channel approximation in the linear regime:

$$\mu = \frac{\partial I_D}{\partial V_G}\bigg|_{V_{SD}} \times \frac{L}{W} \times \frac{t_{ox}}{\varepsilon ox \times V_{SD}}$$

where $\left.\frac{\partial I_D}{\partial V_G}\right|_{V_{SD}}$ is the slope of the curve drain current vs. gate voltage, $L$ is the length of the channel, $W$ is the width of the channel, $t_{ox}$ & $\varepsilon_{ox}$ is the thickness and permittivity of the oxide layer and $V_{sd}$ is the source drain voltage applied ($\leq 5$ V).

The carrier concentration (n) was measured as:

$$n = \frac{\sigma}{e \times \mu}$$

where $\sigma$ is the conductivity, $e$ is an elementary charge and $\mu$ is mobility of the charge carrier.

**Photocurrent Measurements:**

Photocurrent measurements were performed in a cryogenic probe station *CRX-6.5K* (*Lake Shore Desert*) at room temperature and a pressure of $5 \cdot 10^{-6}$ mbar. For this, the film on the bottom gate-bottom contact transistor substrate was transferred into the probe station under minimal exposure to air. The samples were contacted in a two-point probe fashion. Data were recorded using a *2634B SYSTEM SourceMeter* from *Keithley Instruments* operated by the software *test script processor (TSP) express*. As an excitation source, single mode fiber-pigtailed laser diodes from *Thorlabs* operated by a compact laser diode controller *CLD1010* by *Thorlabs* were used: A 637 nm laser diode with a maximal output power of 70 mW and a 847 nm laser diode with a maximal output power of 30 mW. Losses to this theoretical optical power output due to scattering, inefficient coupling into the optical fiber, decollimation of the beam etc. were determined by a calibration sample and an optical power meter to obtain the total incident optical power at the sample surface.

**Transient absorption measurements:**

Transient absorption of thin films of CdSe/I⁻/Zn4APc NCs were studied by broadband pump-probe spectroscopy in a set-up described previously and briefly discussed here.[3,4] The samples were drop-casted on quartz plates (Eso Optics) yielding thin optically dense films of varying thickness. 180 fs laser pulses, generated in a Yb:KGW oscillator at 1028 nm, are split-off to generate a pump and a probe beam. The pump beam energy was varied by nonlinear frequency mixing in an optical parametric amplifier (OPA) and second harmonics generation (Light Conversion, Orpheus). A broadband probe spectrum was generated by focusing the 1028 nm laser light onto a sapphire (500-1500 nm) or a $CaF_2$ (400-600 nm) crystal by nonlinear processes. The probe pulse can be delayed up to 3 ns by an automated delay stage. The majority of the 1028 nm fundamental laser beam is used as the pump pulse for photoexciting the sample (wavelengths 310-1500 nm) after nonlinear frequency mixing in an optical parametric amplifier (OPA) and second harmonics module (Light Conversion, Orpheus). The pump and the probe pulse overlap at the sample position in an ~8° angle. The pump pulse is dumped after the photoexcitation of the sample, while the probe light is led to a detector fiber suitable for the probe spectrum selected (Helios, Ultrafast Systems).

**Time resolved photocurrent measurements:**

Time resolved photocurrent measurements were performed at room temperature under vacuum ($1.5 \cdot 10^{-5}$ mbar) and the devices were kept under vacuum for at least 2 h before starting any measurements. Pulse response measurements were performed using a picosecond pulsed laser driver (Taiko PDL M1, PicoQuant) together with laser heads for 636 nm (pulse length < 500 ps) and 779 nm (pulse length < 500 ps) operation. 3 MHz was selected as repetition rate with an output power of approximately 600 µW. For the step function measurements, a nanosecond diode laser driver (FSL500, PicoQuant) with a laser rise time < 0.5 ns in conjunction with a 635 nm laser diode operated at 3 MHz with a pulse width of 100 ns and an average output power of ≤ 12 mW was used. This laser power is subject to further losses due to scattering, inefficient coupling into the optical fiber, decollimation of the beam etc. The current was preamplified with a FEMTO HSA-Y-1-60 1 GHz high speed amplifier and measured with a Zurich Instruments UHFLI Lock-In Amplifier with Boxcar Averager Function which averages the signal from 33.6M samples. The signals were background corrected. The time resolution was limited to 600 MHz due to the signal input of the Lock-In Amplifier.

**Steady-State photoluminescence measurements:**

Steady-State photoluminescence measurements were performed using an inverted confocal microscope equipped with oil-immersion microscope objective (NA=1.25). The samples were placed in ambient atmosphere and excited using 488nm (optical power on sample 1 mW) laser diode manufactured by *TOPTICA iBEAM-SMART-488*. The photoluminescence data were recorded using an *Acton SpectraPro2300i* spectrometer at -40 °C detector temperature and 4 s acquisition time.

**Time-resolved photoluminescence decay measurements:**

The fluorescence lifetime of CdSe/I$^-$ and CdSe/I$^-$/Zn4APc were measured using a home built scanning confocal microscope.[5] A 488 nm pulse laser (200 µW, 20 MHz) illuminated the sample using a high numerical aperture (NA=1.46) oil objective. The fluorescence was collected by the same objective and then sent via a beam splitter (BS) to a single photon avalanche photodiode (APD) which was connected to a Time-correlated Single Photon Counting Detector (TCSP, HydraHarp 400, **Figure S4**). Decay curves were fitted and analysed by SymPhoTime 64.

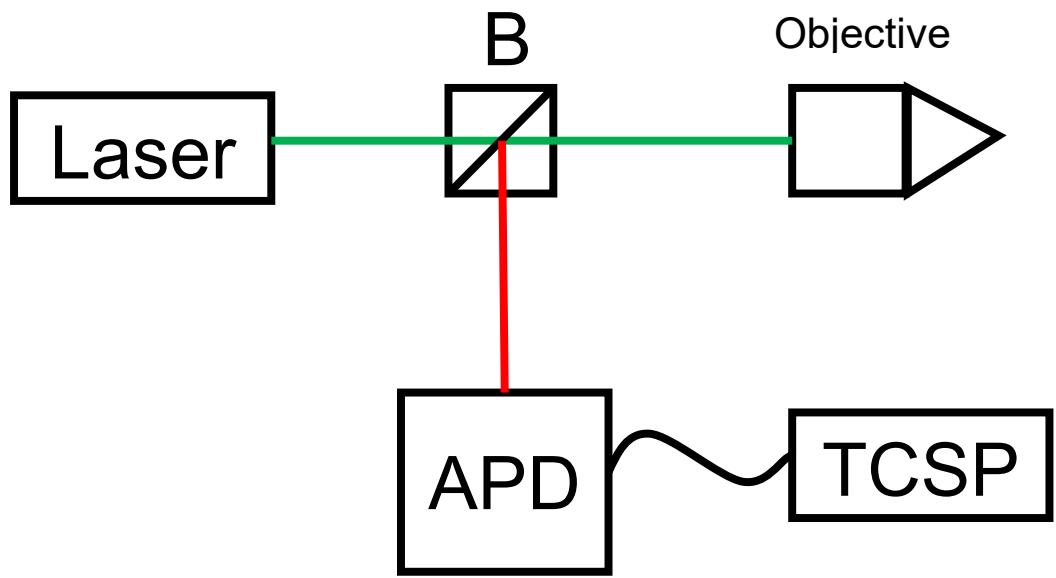

**Fig. S4.** Schematic showing the components of the Time-resolved photoluminescence instrument.